\documentclass[letterpaper,aps,twocolumn,superscriptaddress]{revtex4}
\usepackage{graphicx}
\usepackage{amsmath,amssymb,latexsym}
\usepackage{color}

\def\lsim{\mathrel{\rlap{\lower4pt\hbox{\hskip1pt$\sim$}}
    \raise1pt\hbox{$<$}}}

\begin{document}

\title{UHECR Composition Models}

\author{Andrew~M.~Taylor}
\affiliation{Dublin Institute for Advanced Studies, 31 Fitzwilliam Place, Dublin 2, Ireland\\
Phone: +353 16621333 ext.337, Email: taylora@cp.dias.ie}

\begin{abstract}

In light of the increasingly heavy UHECR composition at the highest energies,
as observed by the Pierre Auger Observatory, the implications of 
these results on the actual source composition and spectra are investigated. 
Depending on the maximum energy of the particles accelerated, sources producing 
hard spectra and/or containing a considerably enhanced heavy 
component appear a necessary requirement. Consideration is made of two archetypal 
models compatible with these results. The secondary signatures expected
, following the propagation of the nuclear species from source to Earth,
are determined for these two example cases.
Finally, the effect introduced by the presence of nG extragalactic magnetic fields 
in collaboration with a large ($80$~Mpc) distance to the nearest source is discussed.

\end{abstract}

\newcounter{pub}

\maketitle

\section{Introduction}
\label{intro}

During the last decade the field of UHECR research has undergone considerable developments
with the completion of extremely large detector facilities. The data from these instruments
has lead to a notable improvement in both the quantity and quality of UHECR measurements.
Following the digestion of this new information, a revision of the UHECR model working hypothesis 
may be due. In particular, measurements sensitive to the UHECR composition have improved 
dramatically with a coherent picture starting to emerge from the ensemble of different 
composition sensitive measurements the Pierre Auger Observatory (PAO) has made \cite{PierreAuger:2011aa}. 
It should be noted that this picture is obscured somewhat when additional observational data from 
the TA experiment are included. The statistical significance of this disagreement, however, 
remains unclear. 
In this study, such additional observational data sets are neglected.

\section{Monte Carlo Modeling}
\label{modelling}

\begin{figure}[h!]
\begin{center}
{\includegraphics[angle=0,width=0.95\linewidth,type=pdf,ext=.pdf,read=.pdf]{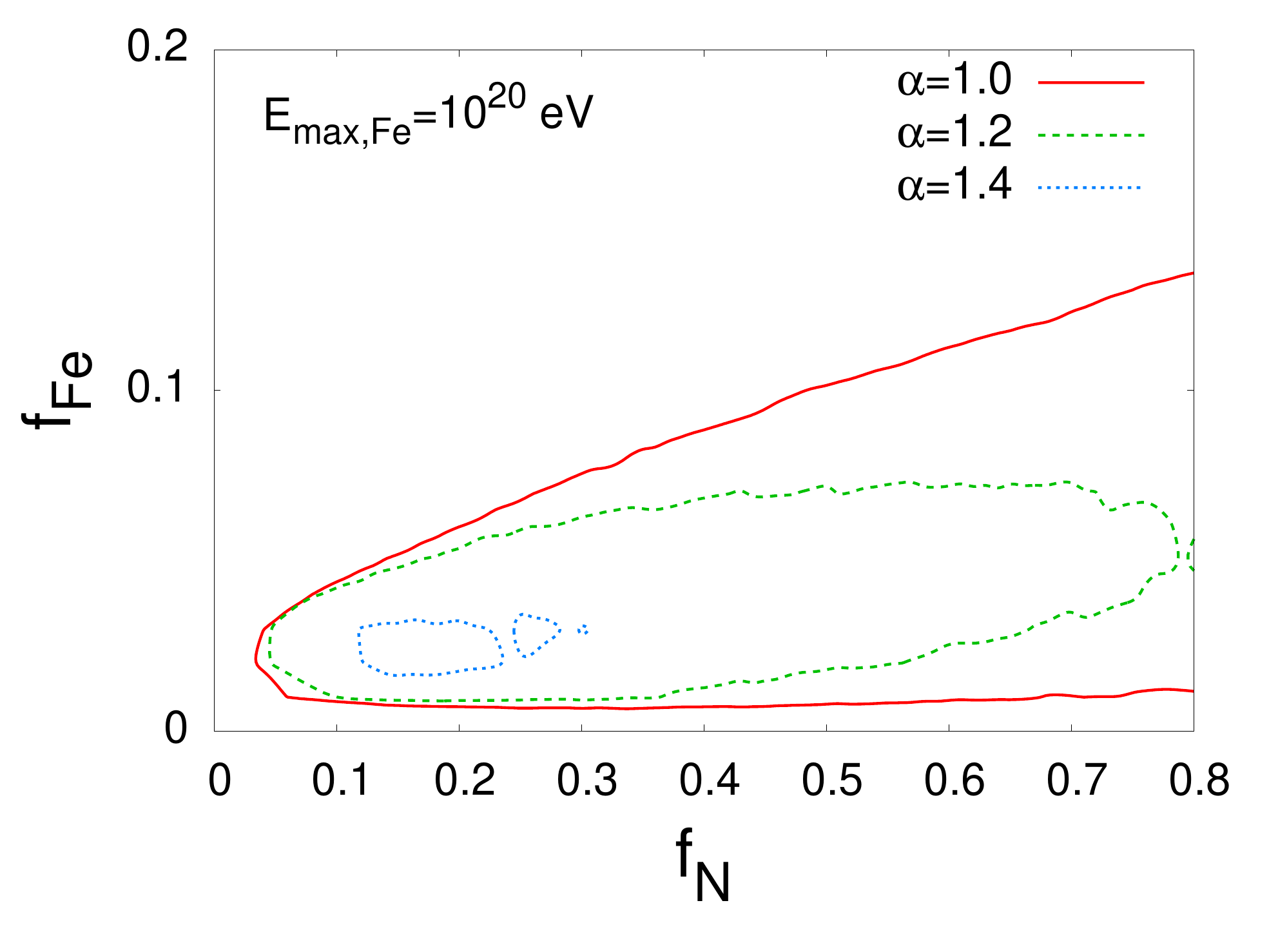}
\includegraphics[angle=0,width=0.95\linewidth,type=pdf,ext=.pdf,read=.pdf]{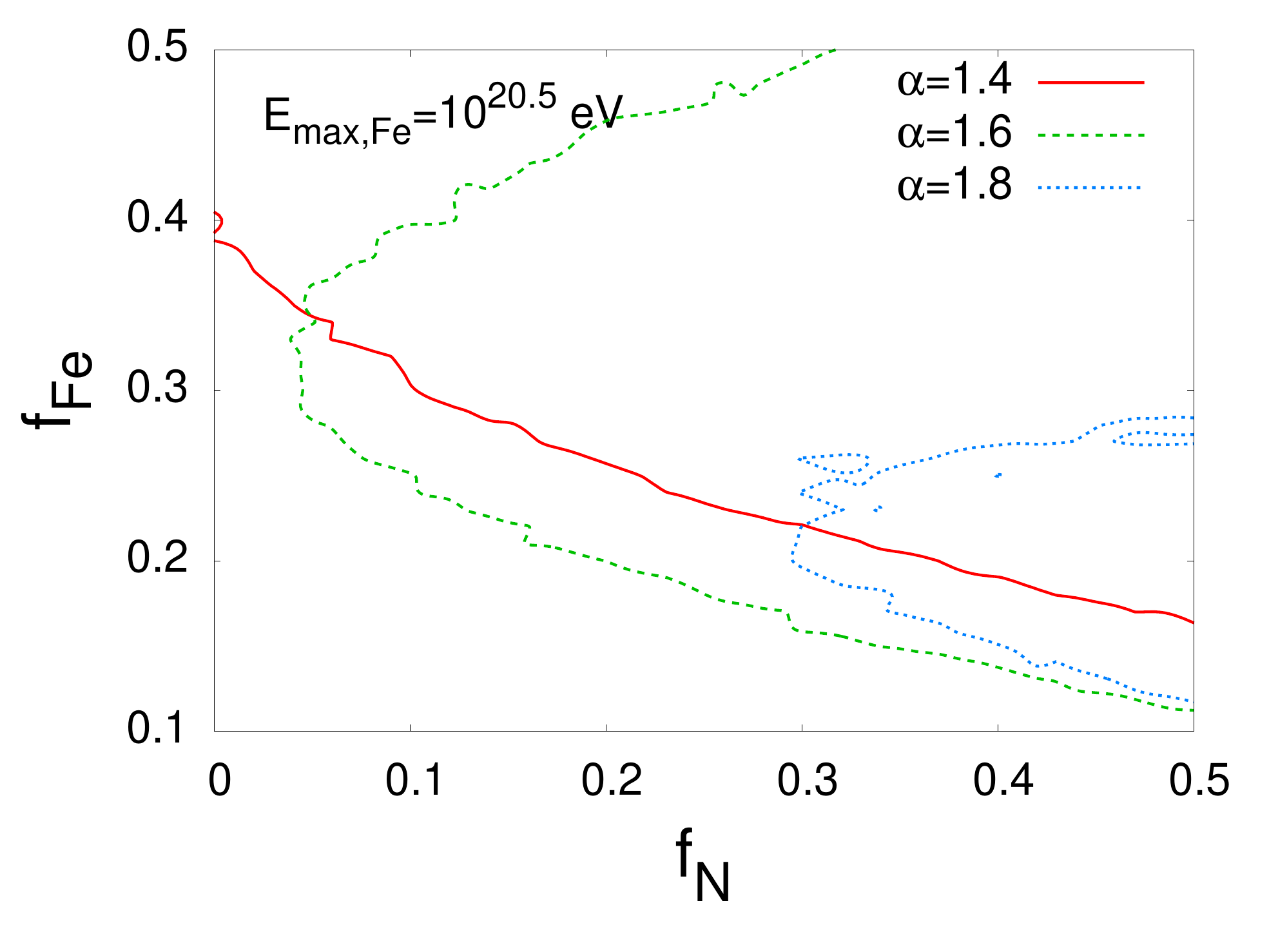}}
\caption{The 95\% good-fit region found following a scan over different iron and nitrogen fractions ($f_{\rm Fe}$ and $f_{\rm N}$), for various spectral index  values. The upper and lower panels show results for $E_{\rm max, Fe}=10^{20}$~eV and $E_{\rm max, Fe}=10^{20.5}$~eV, respectively.}
\label{Parameter_Scan}
\end{center}
\end{figure}

In order to test different hypothesis models, a Monte Carlo description of UHECR propagation is
used, as first described in \cite{Hooper:2006tn}. In this description, UHECR protons and nuclei 
are propagated through the cosmic microwave background (CMB) and cosmic infrared background (CIB) 
radiation fields, undergoing photo-disintegration, photo-pion, pair production , and redshift 
losses as they do so. Though the cross-sections and target photon spectral distributions relevant 
for the proton related energy loss processes are well understood, some uncertainty still remains 
in both the photo-disintegration cross-sections and the CIB spectral distribution relevant for
nuclei propagation. In the present 
study, the description of these adopted are \cite{TALYS} and \cite{Franceschini:2008tp} for the 
cross-sections and CIB spectral distribution respectively.
In sections~\ref{models}, \ref{distinguish}, and \ref{secondaries}, extragalactic magnetic field
(EGMF) effects are neglected. The effects introduced by such fields on the main results are discussed 
in section \ref{distribution}. In order to take account of EGMF effects, the ``delta-approximation'' 
prescription provided in \cite{Taylor:2011ta} is implemented.

To perform a comparison with the PAO measurements, the predicted values of the composition 
sensitive shower profile parameters, $X_{\rm max}$ and $\mathrm{RMS}(X_{\rm max})$, were determined
for each model. In order to encapsulate the uncertainty in the hadronic model description for
these values, the spread in predicted values from four different models 
\cite{qgsjet_11,qgsjet,sibyll,epos} was determined.

The Monte Carlo description was applied to an ensemble of distributed sources whose redshift 
evolution scaled as $(1+z)^{m}$, with $m=3$ from $z_{\rm min}$ (with corresponding nearest source 
distance $L_{\rm min}$) up to $z_{\rm max}=1.5$. An energy spectrum output by each source, of the form
$dN/dE\propto E^{-\alpha}e^{-(E/E_{\rm max, Z})}$, with $E_{\rm max, Z}=(Z/26)E_{\rm max, Fe}$,  was adopted \footnote{Such a
parameterisation should be considered to provide ``effective parameters'', which broadly encapsulate the 
true emission spectrum properties.}. 
Source spectral indices in the range $1<\alpha<3$, and 3-component compositions were scanned over
for both the cutoff energy cases of $E_{\rm max, Fe}=10^{20}$~eV and $E_{\rm max, Fe}=10^{20.5}$~eV. 
Only spectral and composition data points with energies above $10^{18.6}$~eV were used in the analysis. 
The systematic errors for the energy resolution, $X_{\rm max}$, and $\mathrm{RMS}(X_{\rm max})$, were also 
included in the $\chi^{2}$ determination.
The regions of parameter space for which good fits to both the spectral and composition data were found 
are shown in fig.~\ref{Parameter_Scan}. From each of the two cutoff energy results, two example models 
were adopted, whose general characteristics are discussed in the following section.

\section{Composition Models}
\label{models}

\begin{figure}[h!]
\begin{center}
{\includegraphics[angle=0,width=0.9\linewidth,type=pdf,ext=.pdf,read=.pdf]{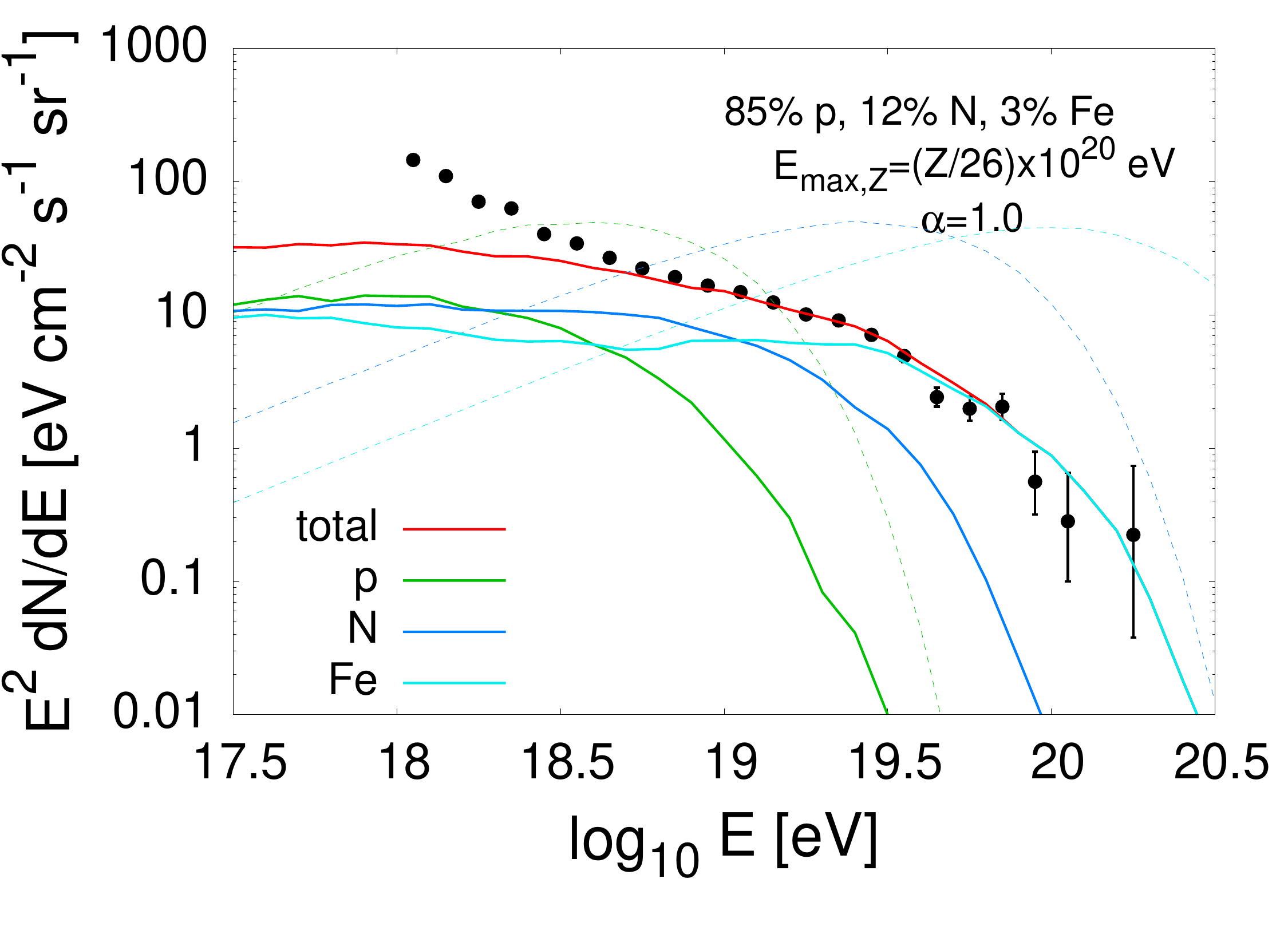}
\includegraphics[angle=0,width=0.9\linewidth,type=pdf,ext=.pdf,read=.pdf]{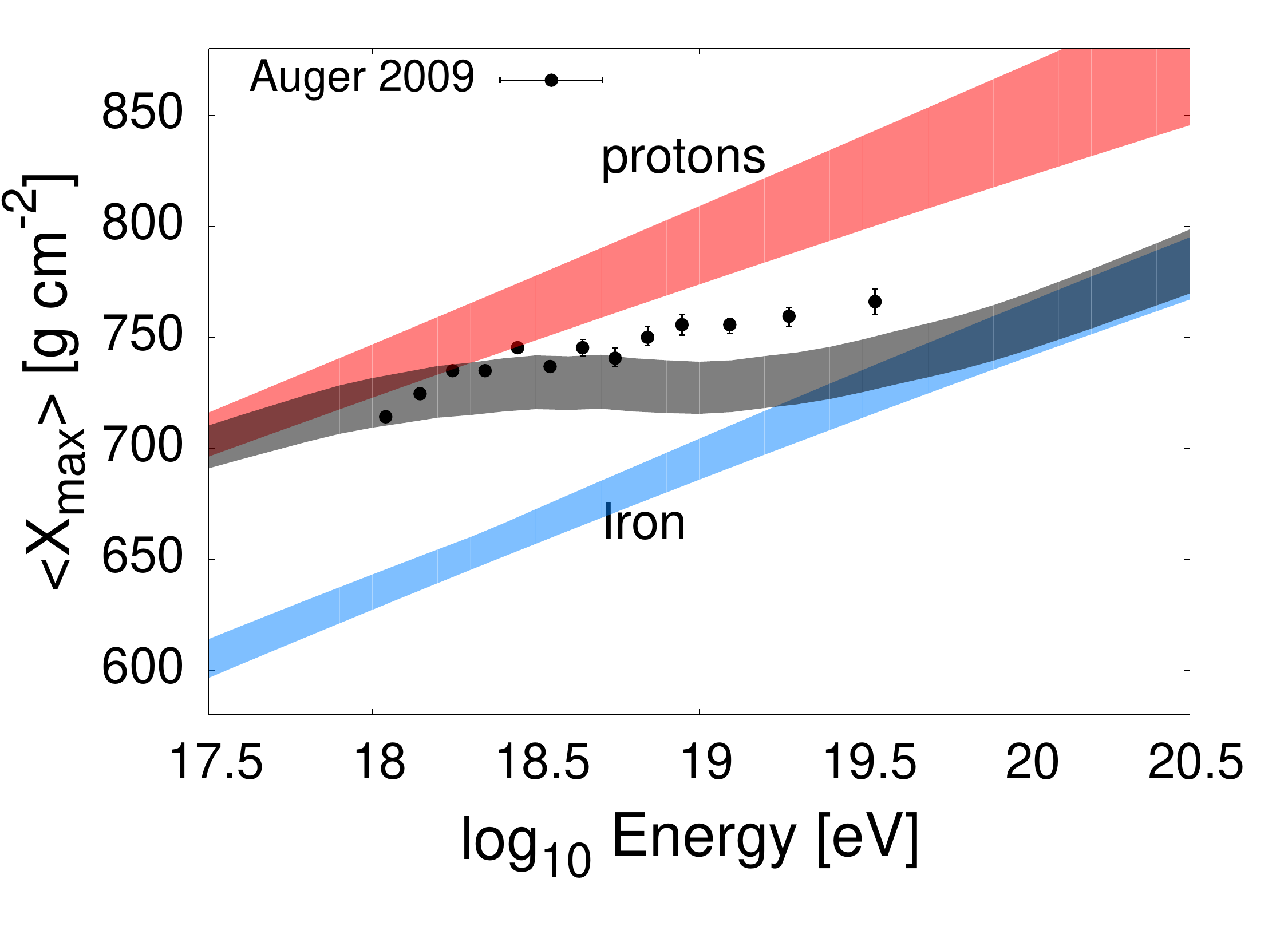}
\includegraphics[angle=0,width=0.9\linewidth,type=pdf,ext=.pdf,read=.pdf]{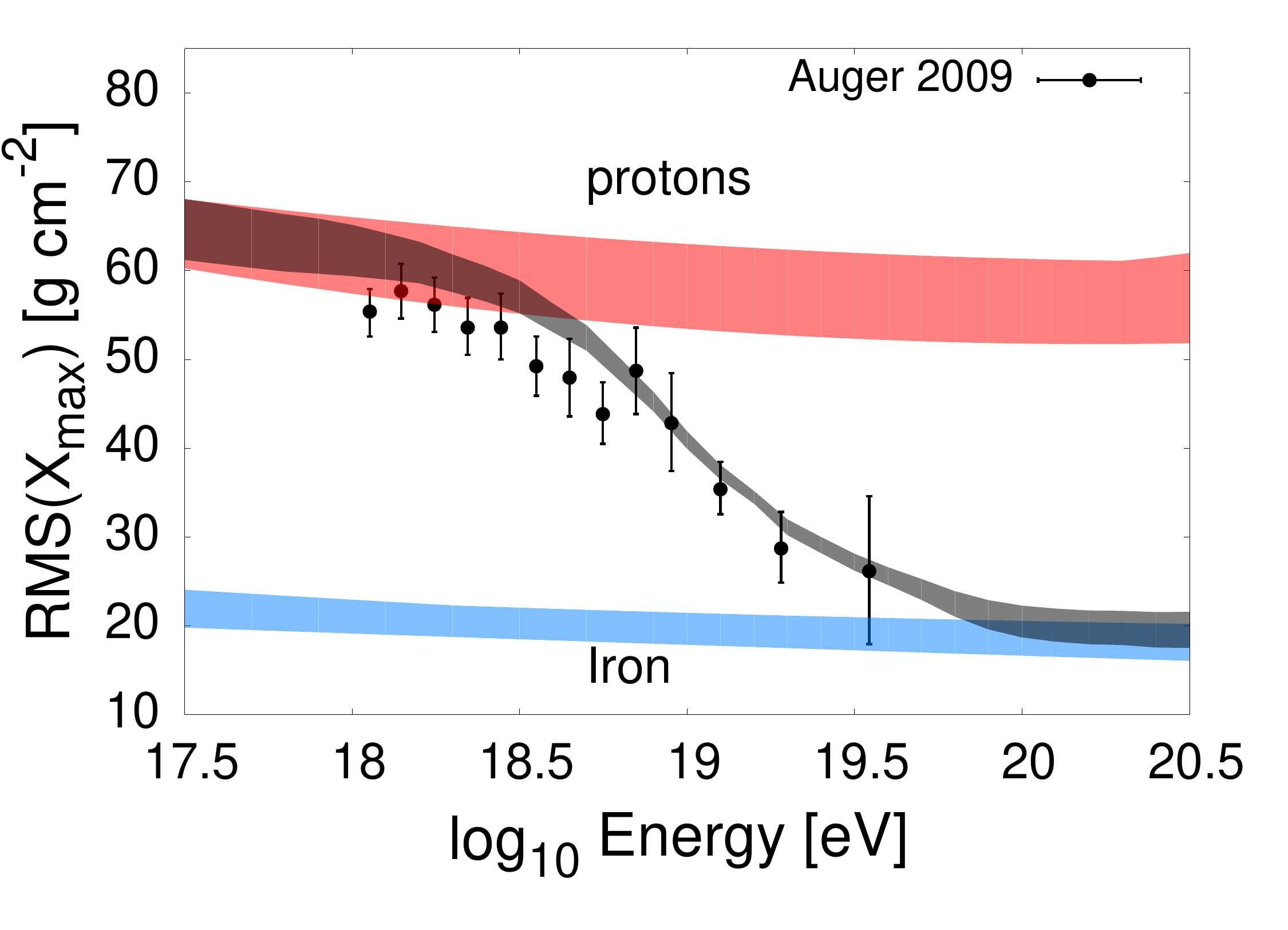}}
\caption{An archetypal example from the first model set. Note a source spectrum of $dN/dE\propto E^{-\alpha}e^{-(E/E_{\rm max, Z})}$, where $E_{\rm max, Z}=(Z/26)E_{\rm max, Fe}$, was adopted. The dashed lines in the top panel show the corresponding spectrum neglecting energy loss processes.}
\label{Model_One}
\end{center}
\end{figure}

\begin{figure}[h!]
\begin{center}
{\includegraphics[angle=0,width=0.9\linewidth,type=pdf,ext=.pdf,read=.pdf]{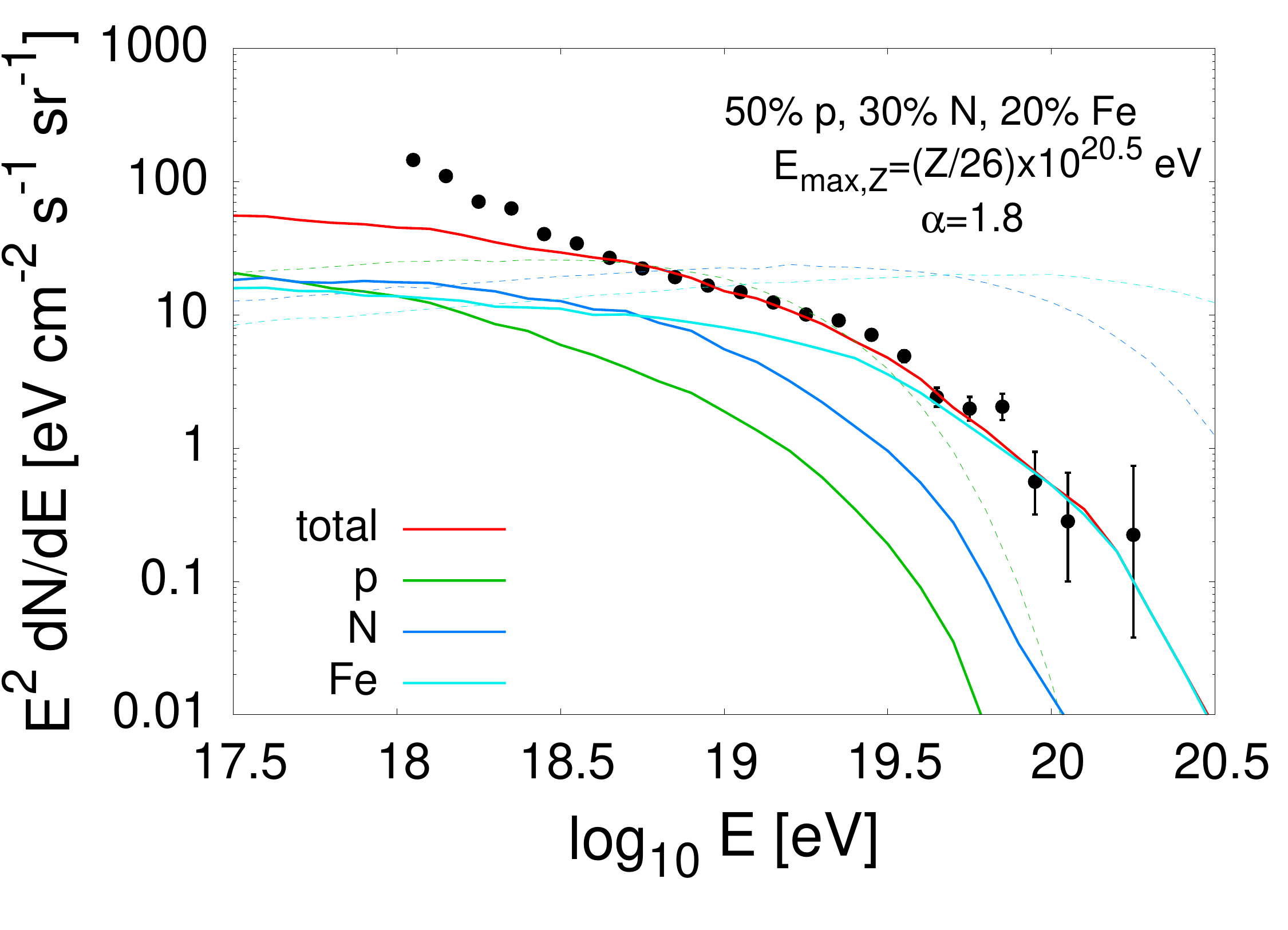}
\includegraphics[angle=0,width=0.9\linewidth,type=pdf,ext=.pdf,read=.pdf]{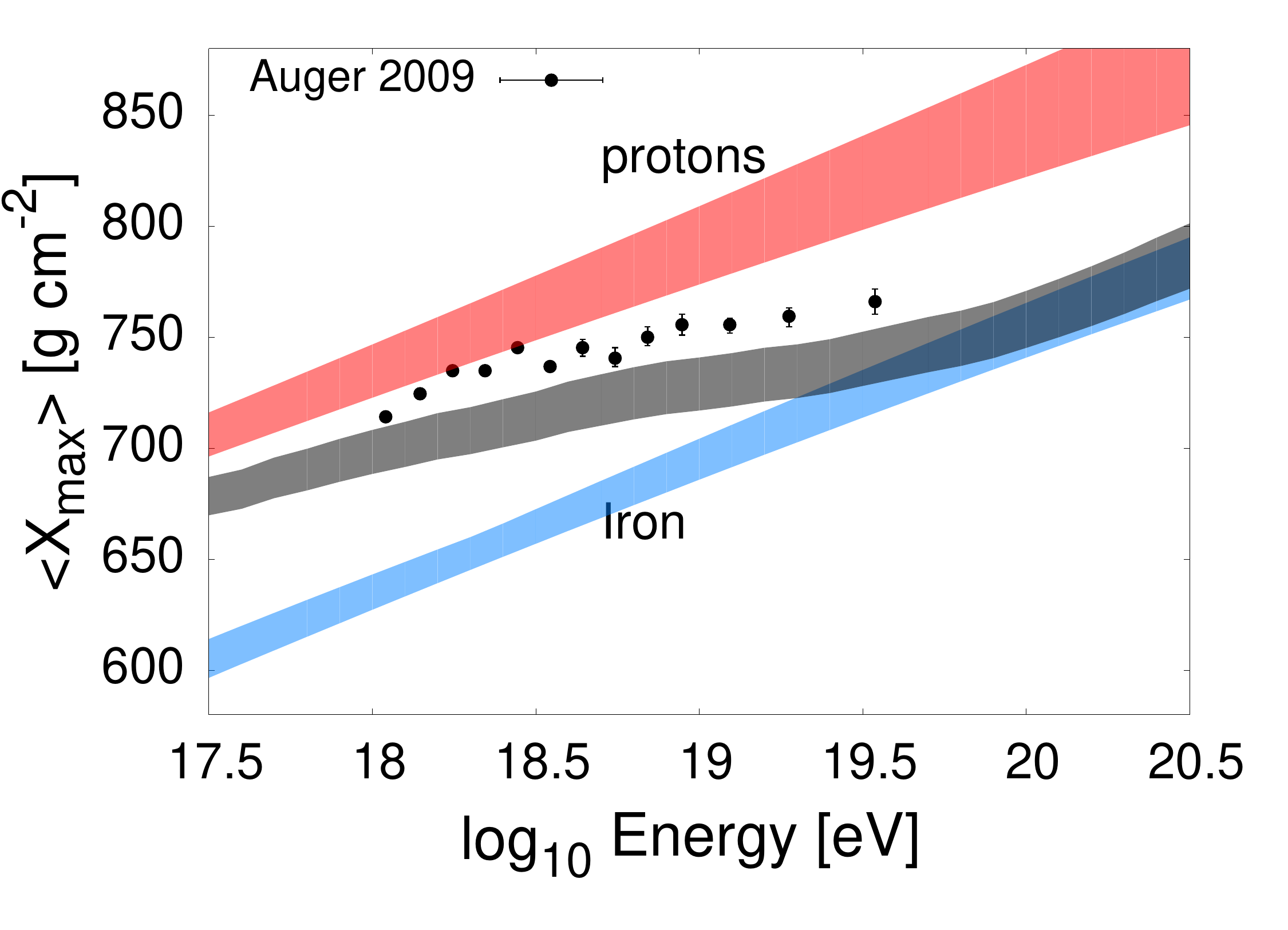}
\includegraphics[angle=0,width=0.9\linewidth,type=pdf,ext=.pdf,read=.pdf]{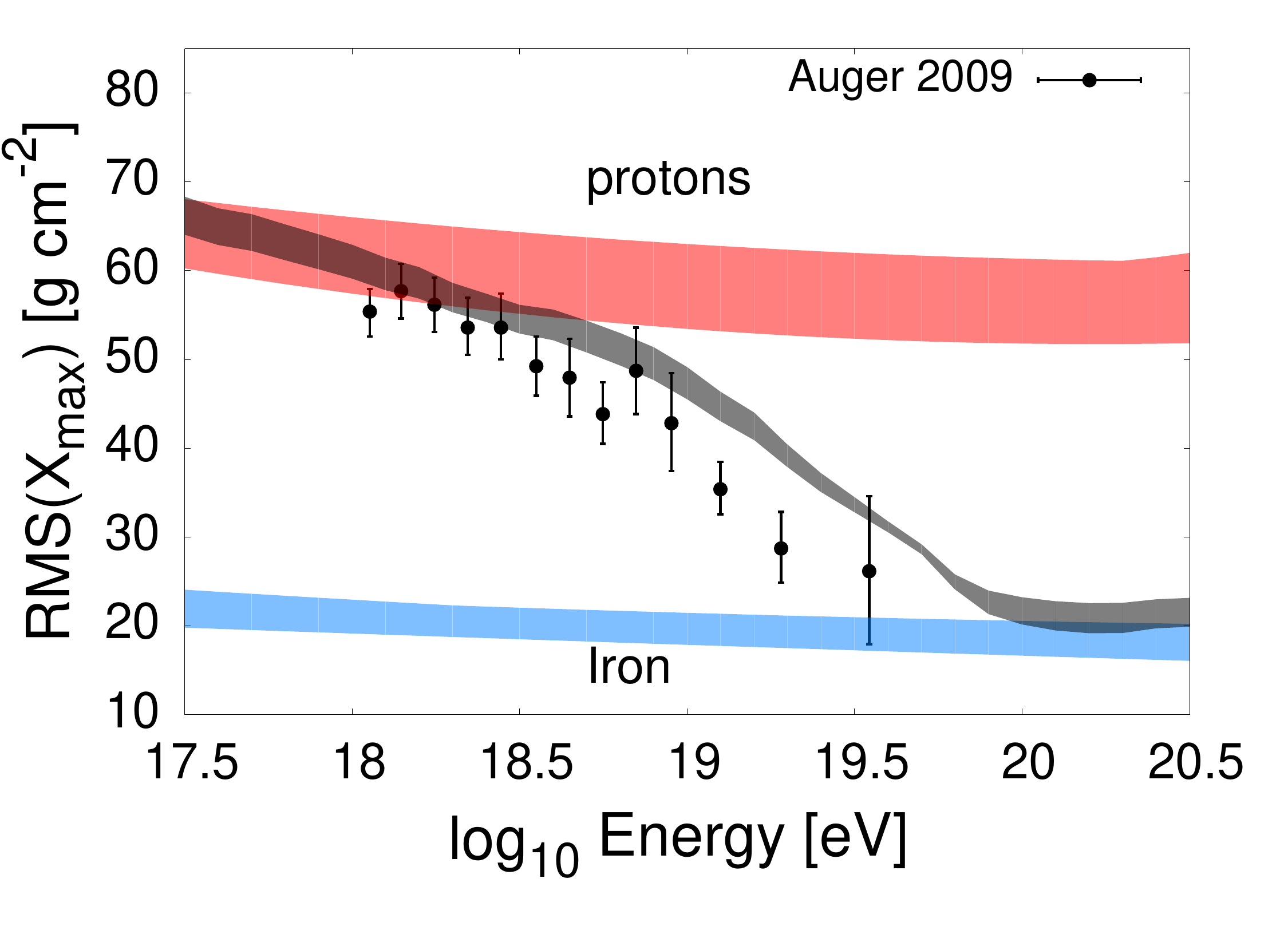}}
\caption{Similar to fig.~\ref{Model_One} for an archetypal example from the second model set.}
\label{Model_Two}
\end{center}
\end{figure}

Applying the Monte Carlo description for CR propagation described in the previous section,
the model parameters required to provide fits to both the PAO composition 
\cite{Abraham:2010yv,Unger:2011ry} and spectral \cite{Abraham:2009wk} data was explored.
In an effort to keep the free parameters of the model to a minimum, only a three
component (light, intermediate, and heavy) admixture  of species is considered.
The composition resolution available with this description is sufficient for the
general purposes of this work.
The landscape of models compatible with recent
PAO observations can be understood by models at the extremities of the distribution.
Broadly, the landscape may be separated into two groups. Archetypal models showing
representative members of these groups are shown in fig.s~\ref{Model_One} and \ref{Model_Two}.
In the spectral plot figures (upper panels), the percentages of the 3 component species 
injected at the source are indicated.

\subsection{Group One- Hard Spectra}

In the first of these model sets, a low cutoff energy and a composition dominated by a light 
component is adopted, with intermediate nuclei taking an abundance at the 
$\sim 10$\% level and heavy nuclei at the $\sim 1$\% level.
An archetypal example of a 
member of this group is shown in fig~\ref{Model_One}. In order to accommodate such a wide spread 
(in terms of mass number $A$) admixture of nuclear species, an extremely hard injection spectrum 
appears necessary.

This hard spectrum requirement originates from the role of the flux from the injected
intermediate mass component in this model, which dominates at energies in the range 
$10^{18.3}$-$10^{19}$~eV, as shown in the top panel of fig.~\ref{Model_One}. The onset of the 
``GZK'' effect \cite{Greisen:1966jv, Zatsepin:1966jv} for such species occurs at lower energies 
than that of iron and protons. Subsequently, the associated steepening of this component also 
occurs at lower energies. Thus, a harder injection spectrum is required in order to counteract 
this effect.

It is worth highlighting that for softer spectra (ie. larger $\alpha$) with $E_{\rm max, Fe}=10^{20}$~eV
fail to provide a reasonable spectral fit. It has been suggested that magnetic fields may
lead to a natural hardening of the spectra \cite{Aloisio:2009sj,Mollerach:2013dza}. 
The realisation of such a scenario, though possible, requires both sources closer than 
$\sim 80$~Mpc and strong EGMF values to be present. Specifically, field strengths stronger 
than 1~nG and coherence lengths shorter than $1$~Mpc are required.

\subsection{Group Two- Enhanced Nuclei Component}
In the second model set, a higher cutoff energy and a significantly enhanced heavy nuclei 
component, relative to the values in the first model set, are adopted. More specifically,
source nuclear component ratios of 50\% proton, 30\% nitrogen, and 20\% iron were selected.
Examples of this model can be found in ref.s~\cite{Hooper:2009fd} and \cite{Allard:2011aa}.

Generally, the higher the iron fraction, the larger the required maximum energy. However,
only a mild increase in $E_{\rm max, Fe}$ is in fact necessary to notably increase the overall 
amount of photo-disintegration undergone by nuclei and lighten the composition at lower 
energies. Similarly, an increase in the iron fraction allows 
a decrease in the injection spectral index. This follows from the increased stability
of UHE heavy nuclei relative to lighter nuclei at the same energy due to their reduced
Lorentz factor.

In the extreme example case of this group, as can be appreciated from a consideration of
the top panel in fig.~1 of \cite{Taylor:2011ta}, the source composition could 
conceivably constitute of purely iron nuclei, for which case a further increase in the maximum 
energy (ie. to $E_{\rm max, Fe}=10^{21}$~eV) is able to produce a reasonable fit to full 
spectral and compositional data set.

With the hard spectra and heavy nuclei component fits not being mutually exclusive,
the possibility also exists for models which adopt both the hard spectra and a heavy nuclei 
component features from the two groups. Indeed, recently a 
physical realisation of such a model was proposed \cite{Fang:2013cba}.

\section{Similarities and Differences of Groups}
\label{distinguish}

Observationally, the two distinguished groups are presently not easily separable.
At ``low'' energies, both groups require a separate, softer, component to
take over below an energy of $10^{18.5}$~eV. Historically, such a component has been referred
to as the Galactic cosmic ray flux \cite{Hill:1985,Wibig:2004ye}. However, it should be
noted that the expected change in composition for the ``ankle'' Galactic-Extragalactic
source transition scenarios is the reverse of that actually observed \cite{Allard:2005cx},
with the composition undergoing a transition, apparently becoming heavier with energy
above this spectral feature.
Furthermore, the expected anisotropy level for such a high transition energy 
\cite{Giacinti:2011ww} may already violate present dipole anisotropy limits 
\cite{Abreu:2012lva}. These limits, therefore, may already indicate that the transition 
from Galactic to extragalactic sources occurs at energies below the ankle. A
lightening in the composition associated with this transition would also 
fit into such a scenario \cite{Apel:2013ura}.
At the highest observed energies ($>10^{20}$~eV), the flux in both cases is dominated 
by the iron component. Thus, once again the expectations from the two model groups at 
these energies are quite similar.

Though commonalities do exist between these differing viable models, the theoretical
challenges that they place on the source environment are quite different.
On the one hand, the first group models require a composition seemingly more compatible
with that observed in lower energy Galactic cosmic rays.
However, accompanying this requirement, the sources are also required to produce
UHECR with extremely hard injection spectra, far harder than that typically produced
by non-relativistic first order Fermi shock acceleration. This said, alternative
acceleration scenarios, such as drift acceleration at relativistic shocks 
\cite{Baring:2010tn}, do exist, and can naturally give rise to hard spectra.
Indeed, the need for hard spectra to be produced by the source demanded by the
current observations, may be revealing something fundamental to us about the acceleration
mechanism at play within the source.

On the other hand, the second group of UHECR source models require spectra more
consistent with the Fermi first order acceleration mechanism. However, for these models, 
considerable enhancement of the heavy nuclei component in the UHECR flux produced by the 
sources is required. Such an enhancement of the heavy nuclear component may also not be 
such a surprise. At ``low'' energies, the abundance ratios of iron and most other heavy 
elements already require an enhancement factor of $\sim 30$ above that measured locally 
within the solar system \cite{Drury:2012md,Drury:1999}.

In order to distinguish between these differing scenarios, correlations studies of UHECR 
with specific nearby sources hold great promise. Indeed, such correlations provide the 
possibility for determining the actual source spectrum output by the nearest objects, 
following a similar vein to that already attempted for the case of Cen~A in \cite{Clay:2010id}.
The secondary particles produced during the propagation of the UHECR en-route to
Earth may also shed light on the source origin \cite{Taylor:2009we}. A consideration of 
the secondaries fluxes expected from the two source models is made in the following section.

\section{Secondary Signatures}
\label{secondaries}

Hard cosmic ray spectrum at the source ($\alpha<2$) lead to the injection of the bulk
of the energy flux at the cutoff energy, $E_{\rm max,Z}$.
As indicated by the difference in injected and arrive fluxes in fig.s~\ref{Model_One} and 
\ref{Model_Two}, both sets of models lead in the injection of a notable amount of energy flux 
into losses, primarily through $e^{+}e^{-}$ pair production interactions, with $\pi$ losses
contributing to a lesser extent due to the energy flux being injected below the threshold 
energy for this process. Following the injection 
of these pairs and gamma-rays (produced through $\pi^{0}$ losses) into the inter-galactic 
medium, the subsequent development of electromagnetic
cascades leads to this energy being reprocessed, finally settling in a more stable configuration
in the form of GeV photon flux. Thus applying constraints on the Fermi-LAT gamma-ray background
flux \cite{Abdo:2010nz} subsequent constraints on the accumulated amount of CR losses feeding 
into cascades can be found \cite{Berezinsky:2010xa,Ahlers:2010fw}. The limit on the total 
diffuse flux injected into cascades through pair-losses and pion losses amounts to 
$\sim 1400$~eV~cm$^{-2}$~s$^{-1}$~sr$^{-1}$, with a corresponding energy density of 
$\sim 5.8\times 10^{-7}$~eV~cm$^{-3}$ \cite{Berezinsky:2010xa}.

The calculation of the energy injected into the cascades is tracked using the
Monte Carlo description, with energy losses injected at redshift $z$ diluting as 
$(1+z)^{-4}$ \cite{Ahlers:2010fw}. 

The integrated energy flux injected into cascades for our archetypal ``Group 1'' model  
is $\sim 250$~eV~cm$^{-2}$~s$^{-1}$~sr$^{-1}$ for sources distributed with $m=3$ up to 
$z_{\rm max}=1.5$. The contribution ratios to this flux from the different nuclear species
injected (p:N:Fe) are approximately (0.45:0.30:0.25).
Though this sits below the present 1~year Fermi constraints, 
such a level may be probed by future measurements. 
In comparison, the energy flux injected into cascades for our archetypal ``Group 2'' 
model is $\sim 220~$eV~cm$^{-2}$~s$^{-1}$~sr$^{-1}$, with contribution ratios of 
(0.50:0.30:0.20) from the different nuclear species injected.
These results demonstrate that despite the 
reduced $E_{\rm max, Fe}$ value, the ``Group 1'' model feeds a comparable amount of energy flux into 
cascades compared to our ``Group 2'' model. This fact may be appreciated from the dotted 
lines in fig.s~\ref{Model_One} and \ref{Model_Two}, which represent the injected spectra 
without losses. The considerable energy flux required to be injected by the sources in both 
models highlights the need for the sources to be ``efficient'' with regards not spilling too 
much of the source power into cascade losses. 

Consideration is also worth paying to the corresponding integrated neutrino energy flux generated 
through the decay of $\pi^{+/-}$ losses. For both models this sits at the level of approximately
1~eV~cm$^{-2}$~s$^{-1}$~sr$^{-1}$, with contribution ratios of approximately (0.60:0.25:0.15) 
and (0.70:0.20:0.10) from the different nuclear species injected (p:N:Fe), for the two group
models, respectively. The detection of this flux is also challenging for current neutrino 
telescopes such as IceCube, whose 3 year integrated diffuse flux limit at these energies is 
anticipated to probe levels down to $\sim$7~eV~cm$^{-2}$~s$^{-1}$~sr$^{-1}$ \cite{Abbasi:2011ji}.


It should be noted that the numbers obtained above by no means provide upper/lower bounds
on either flux due to their dependence on the source evolution parameter $m$, set to
3 for the above calculations. Furthermore, with $E_{\rm max,Z}$ values close to the GZK
cutoff energy for the different species considered, the neutrino flux expectations are also 
sensitive to the particular cutoff parameterisation adopted.
Lastly it is worth mentioning that with the dominant component of these calculated
fluxes originating from the population of protons injected, the predicted flux is
sensitive to the light component fraction present in UHECR accelerated by the source.

Though the detection of either of these secondary fluxes is presently challenging, the
GeV gamma-ray or EeV neutrino windows offer great opportunity for directly probing the 
early evolution of UHECR sources.

\section{EGMF and Local Source Distribution Effects}
\label{distribution}

\begin{figure}[h!]
\begin{center}
{\includegraphics[angle=0,width=0.9\linewidth,type=pdf,ext=.pdf,read=.pdf]{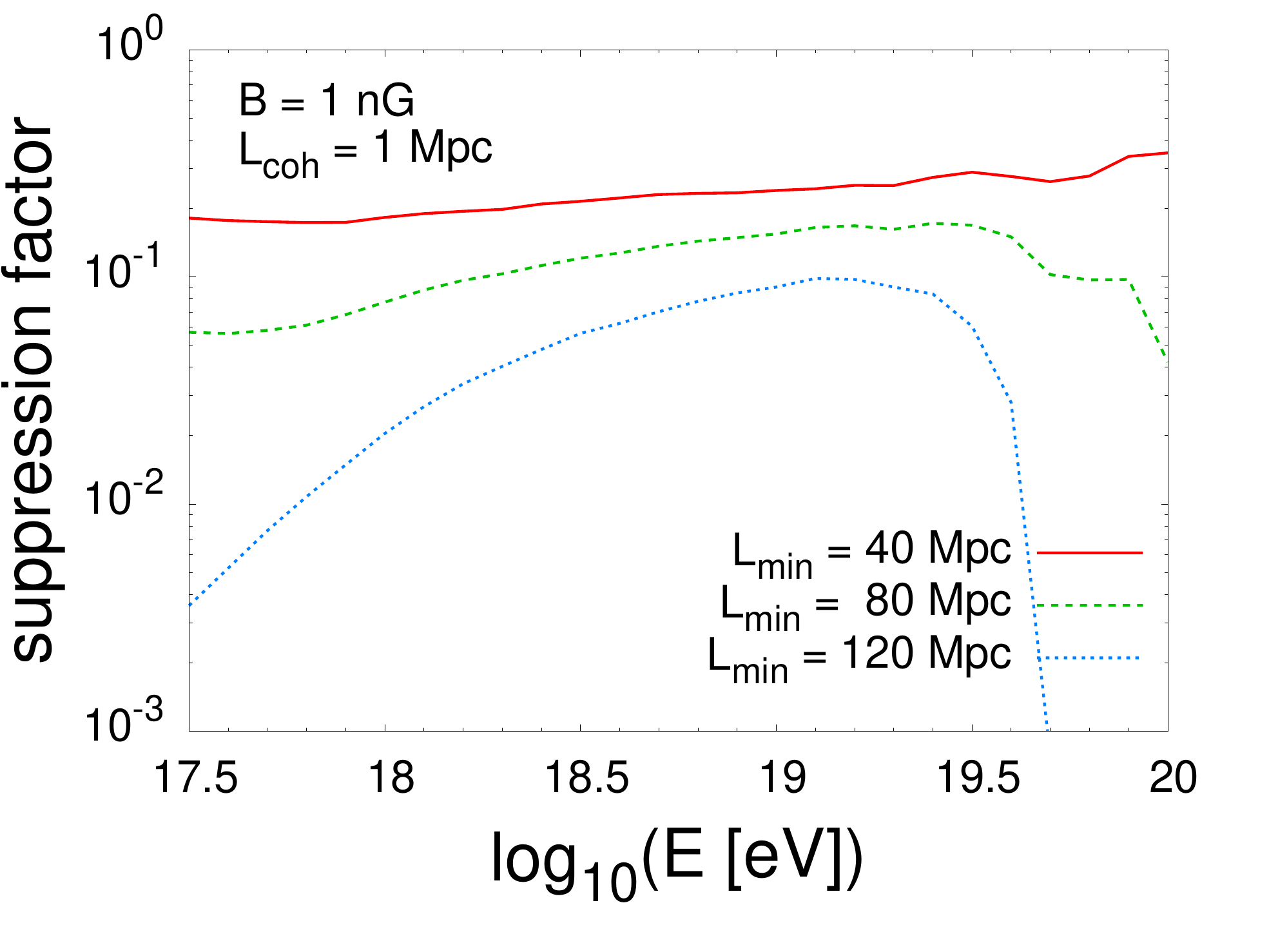}}
\caption{The factor by which the arriving iron flux in the ``Group 2'' model is suppressed, relative to the $0$~G case, due to the presence of an intervening 1~nG EGMF for $L_{\rm min}=40$,~$80$,~$120$~Mpc, with energy loss processes taken into account.}
\label{suppression}
\end{center}
\end{figure}

\begin{figure}[h!]
\begin{center}
{\includegraphics[angle=0,width=0.9\linewidth,type=pdf,ext=.pdf,read=.pdf]{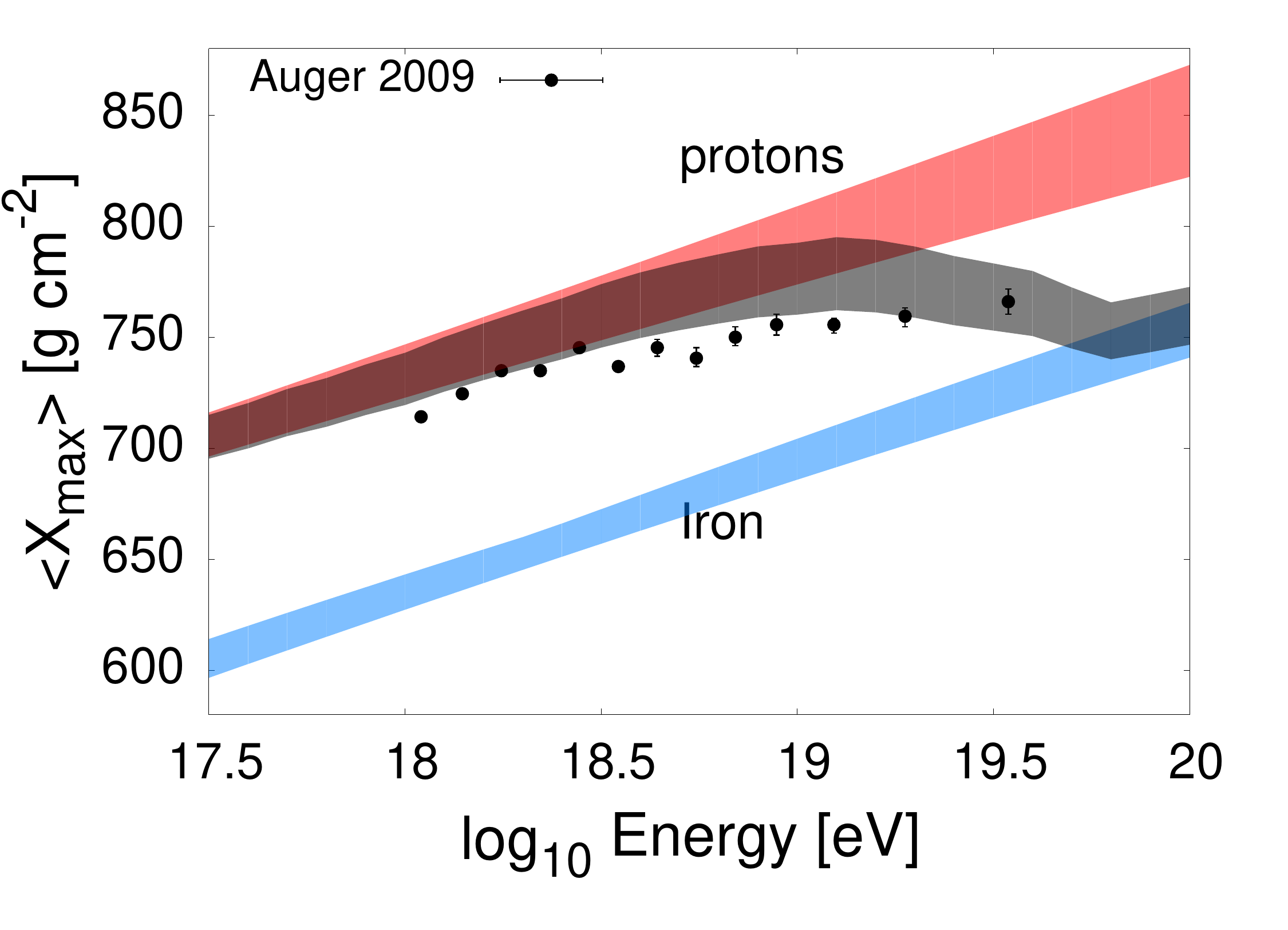}
\includegraphics[angle=0,width=0.9\linewidth,type=pdf,ext=.pdf,read=.pdf]{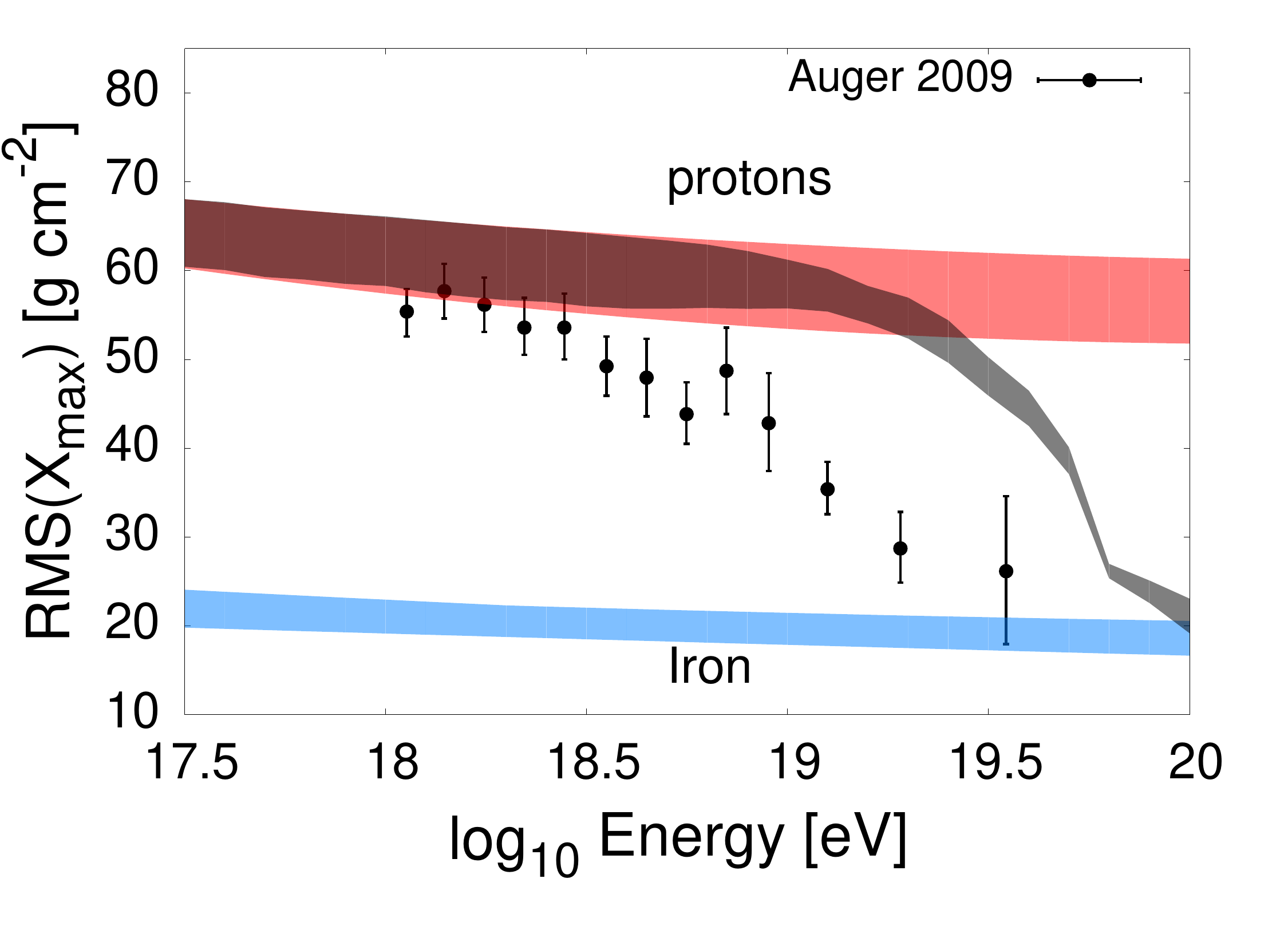}}
\caption{Same as for fig.~\ref{Model_Two} (middle and bottom panel), with the presence of an intervening 1~nG EGMF with $L_{\rm coh}=1~$Mpc added, for the case with $L_{\rm min}=80$~Mpc, with energy loss processes taken into account.}
\label{lightening}
\end{center}
\end{figure}

The UHECR source distribution has thus far been assumed to be locally homogeneous with 
sources existing in the immediate proximity to the Galaxy, down to the propagation length 
scale, the smallest length scale used in the Monte Carlo (ie. 100~kpc). However, as was 
discussed in \cite{Taylor:2011ta}, a more distant nearest source may significantly alter 
the arriving spectrum, particularly at the highest energies observed.

Furthermore, the presence of strong (nG) EGMF has recently been considered as an agent 
to resolve the hard injection spectra
discussed here \cite{Mollerach:2013dza}. Indeed, nG EGMF values
and source distances $\sim$100~Mpc can lead to a hardening of the spectrum as suggested.
However, the associated even harder GZK cutoff expected for such a model \cite{Taylor:2011ta} 
poses an obstacle for such a resolution of the hard injection spectra problem.
The effect on the spectrum introduced by such intervening field strengths coupled with a
``large'' distance to the nearest source is demonstrated explicitly in fig.~\ref{suppression}. 
In this figure, the factor by which the arriving flux is suppressed, relative to the $0$~G EGMF case, 
due to the presence of an intervening nG EGMF is shown. The key point being that
the introduction of a distance to the first source, $L_{\rm min}$, reduces the arriving
flux significantly both at low and high energies. The former of these effects resulting from the 
reduced ability of flux from the nearest source to diffusively propagate to Earth within a Hubble
time. The latter effect, on the other hand, resulting from the energy loss lengths becoming smaller or 
comparable to the distance of the first source.

Furthermore, as demonstrated explicitly for the ``Group 2'' model in fig.~\ref{lightening},
the introduction of a large $L_{\rm min }$ value (ie. $\sim $80~Mpc) in
collaboration with a strong nG EGMF can lead to a significant increase in the
amount of photo-disintegrated nuclei arriving at Earth. The consequence of this is a 
lightening of the arriving composition, which can lead to conflict with the PAO $X_{\rm max}$ and 
$\mathrm{RMS}(X_{\rm max})$ measurements.

Such composition related issues are somewhat alleviated by the reduction of the cutoff energy down
to $E_{\rm max, Fe}=10^{20}$~eV. Such a decrease in the cutoff energy reduces the contamination of 
photo-disintegrated products in the arriving flux, improving both the $X_{\rm max}$ and 
$\mathrm{RMS}(X_{\rm max})$ fits to the data. 
The spectral related issues, however, remain unresolved 
by such a alteration of the cutoff energy. 
Indeed, a magnetic horizon solution to the hard spectra problem requires EGMF strengths 
stronger than nG with coherence lengths shorter than 1~Mpc to be considered 
\cite{Mollerach:2013dza}. Such values sit close to present EGMF upper limits 
\cite{Kronberg:1976,Kronberg:1982,Blasi:1999hu}.

Lastly, it should be borne in mind that on top of the effects of the nearest source considered, 
the local realisation of the distribution of sources, will also bear their imprint on the arriving 
UHECR spectrum \cite{Ahlers:2012az}. Current limits on the density of the sources sit at the level 
$10^{-6}-10^{-5}$~Mpc$^{-3}$ \cite{Takami:2012uw,Kashti:2008bw}. With these limits dictating the maximal 
size of ``cosmic variance'' effects, the current constraints leave open a considerable
range of uncertainty. Thus, the actual realisation of the local source distribution can 
potentially also have significant effects on spectral and composition fits.

\section{Conclusion}
\label{conclusion}

Recent improvements in UHECR detectors are now providing spectral and composition
data with considerably reduced (statistical) errors relative to that previously
obtainable. Though still early days in this new era of UHE astrophysics, consideration
is spent here on what recent data may be suggesting with regards UHECR models. In order
to dissect the landscape of possible models consistent with recent results, two 
distinct archetypal examples are selected.

Depending on the maximum energy of particles accelerated by the source, 
both hard spectra and enhanced nuclear component model groups are found to give
reasonable fits to the spectrum and composition sensitive measurements made
by the PAO. The arriving flux and composition for two archetypal cases from
these model groups are provided in fig.s~\ref{Model_One} and \ref{Model_Two}.
These model results encapsulate the key attributes 
from the broad range of models considered, with each in turn placing differing 
requirements on the UHECR source and its environment.

The secondary loss signals expected from both these models are demonstrated to be difficult
to detect with current GeV $\gamma$-ray and EeV neutrino detectors.

Finally, the effect introduced by the presence of nG EGMF with 1~Mpc 
coherence lengths in collaboration with a ``large'' distance to the nearest source are considered. 
Such a strong EGMF value was shown to be able to both harden the arriving spectrum and lighten 
its associated nuclear composition. Though the first of these effects allow for somewhat softer 
injection spectra, the associated harder GZK cutoff feature is problematic. On top of such 
nearest source effects, the actual distribution of the ensemble of local UHECR sources is highlighted
to also be capable of imprinting its effects on the arriving spectrum and composition.

\section*{Acknowledgments}
I thank Felix Aharonian  and Markus Ahlers for useful discussions during the development of this
work. I would also like to thank the anonymous referee for helpful corrections to the text. 
Lastly, I thank Keith Rochford for setting up a new cluster 
at DIAS, which proved invaluable for the numerical evaluation of the random walk results 
presented. I acknowledge a Schroedinger fellowship at DIAS.


\begin{thebibliography}{20}

\bibitem{PierreAuger:2011aa}
  P.~Abreu {\it et al.}  [Pierre Auger Collaboration],
  arXiv:1107.4804 [astro-ph.HE].

\bibitem{Hooper:2006tn}
  D.~Hooper, S.~Sarkar and A.~M.~Taylor,
  Astropart.\ Phys.\  {\bf 27} (2007) 199
  [astro-ph/0608085].

\bibitem{TALYS}
  A.~J.~Koning, S.~Hilaire and M.~C.~Duijvestijn, 
  EDP Sciences, 2008, p.~211-214.

\bibitem{Franceschini:2008tp}
  A.~Franceschini, G.~Rodighiero and M.~Vaccari,
  arXiv:0805.1841 [astro-ph].

\bibitem{Taylor:2011ta}
  A.~M.~Taylor, M.~Ahlers and F.~A.~Aharonian,
  Phys.\ Rev.\ D {\bf 84} (2011) 105007
  [arXiv:1107.2055 [astro-ph.HE]].

\bibitem{qgsjet_11}
  S.S. Ostapchenko, 
  Nucl.\ Phys.\ Proc.\ Suppl.\ {\bf 151} (2006), 143

\bibitem{qgsjet}
  N.N. Kalmykov and S.S.\ Ostapchenko, Phys.\ Atom.\ Nucl.\ {\bf 56} (1993), 346;
  S.S. Ostapchenko, Nucl.\ Phys.\ Proc.\ Suppl.\ {\bf 151} (2006), 143;
  N.~N.~Kalmykov, S.~S.~Ostapchenko and A.~I.~Pavlov,
  Nucl.\ Phys.\ Proc.\ Suppl.\  {\bf 52B}, 17 (1997).

\bibitem{sibyll}
  E.~J.~Ahn, R.~Engel, T.~K.~Gaisser, P.~Lipari and T.~Stanev,
  arXiv:0906.4113 [hep-ph];
  R.~S.~Fletcher, T.~K.~Gaisser, P.~Lipari and T.~Stanev,
  Phys.\ Rev.\  D {\bf 50}, 5710 (1994).

\bibitem{epos}
  T.\ Pierog and K.\ Werner, Phys.\ Rev.\ Lett.\ {\bf 101}
  (2008), 171101;
  K.~Werner and T.~Pierog,
  AIP Conf.\ Proc.\  {\bf 928}, 111 (2007)
  [arXiv:0707.3330 [astro-ph]].

\bibitem{Abraham:2010yv}
  J.~Abraham {\it et al.}  [Pierre Auger Observatory Collaboration],
  Phys.\ Rev.\ Lett.\  {\bf 104} (2010) 091101
  [arXiv:1002.0699 [astro-ph.HE]].

\bibitem{Unger:2011ry}
  M.~Unger and f.~t.~P.~Collaboration,
  arXiv:1103.5857 [astro-ph.HE].

\bibitem{Abraham:2009wk}
  J.~Abraham {\it et al.}  [The Pierre Auger Collaboration],
  arXiv:0906.2189 [Unknown].




\bibitem{Greisen:1966jv}
  K.~Greisen,
  Phys.\ Rev.\ Lett.\  {\bf 16}, 748 (1966).

\bibitem{Zatsepin:1966jv}
  G.~T.~Zatsepin and V.~A.~Kuz'min,
  JETP Lett.\  {\bf 4}, 78 (1966)
  [Pisma Zh.\ Eksp.\ Teor.\ Fiz.\  {\bf 4}, 114 (1966)].

\bibitem{Aloisio:2009sj}
  R.~Aloisio, V.~Berezinsky and A.~Gazizov,
  Astropart.\ Phys.\  {\bf 34} (2011) 620
  [arXiv:0907.5194 [astro-ph.HE]].

\bibitem{Mollerach:2013dza}
  S.~Mollerach and E.~Roulet,
  arXiv:1305.6519 [astro-ph.HE].

\bibitem{Hooper:2009fd}
  D.~Hooper and A.~M.~Taylor,
  Astropart.\ Phys.\  {\bf 33} (2010) 151
  [arXiv:0910.1842 [astro-ph.HE]].

\bibitem{Allard:2011aa}
  D.~Allard,
  Astropart.\ Phys.\  {\bf 39-40} (2012) 33
  [arXiv:1111.3290 [astro-ph.HE]].

\bibitem{Fang:2013cba}
  K.~Fang, K.~Kotera and A.~V.~Olinto,
  JCAP {\bf 1303} (2013) 010
  [arXiv:1302.4482 [astro-ph.HE]].






\bibitem{Wibig:2004ye}
  T.~Wibig and A.~W.~Wolfendale,
  J.\ Phys.\ G {\bf 31} (2005) 255
  [astro-ph/0410624].

\bibitem{Hill:1985}
  C.~T.~Hill and D.~N.~Schramm,
  Phys. Rev. D, Vol. 31, Iss. 3 (1985)

\bibitem{Allard:2005cx}
  D.~Allard, E.~Parizot and A.~V.~Olinto,
  Astropart.\ Phys.\  {\bf 27} (2007) 61
  [astro-ph/0512345].

\bibitem{Giacinti:2011ww}
  G.~Giacinti, M.~Kachelriess, D.~V.~Semikoz and G.~Sigl,
  JCAP {\bf 1207} (2012) 031
  [arXiv:1112.5599 [astro-ph.HE]].

\bibitem{Abreu:2012lva}
  P.~Abreu {\it et al.}  [ Pierre Auger Collaboration],
  ApJL, 762, L {\bf 13} (2012)
  [arXiv:1212.3083 [astro-ph.HE]].

\bibitem{Apel:2013ura}
  W.~D.~Apel, J.~C.~Arteaga-Velàzquez, K.~Bekk, M.~Bertaina, J.~Blümer, H.~Bozdog, I.~M.~Brancus and E.~Cantoni {\it et al.},
  Phys.\  Rev.\  D 87, {\bf 081101} (R) (2013)
  [arXiv:1304.7114 [astro-ph.HE]].

\bibitem{Baring:2010tn}
  M.~G.~Baring,
  arXiv:1002.3848 [astro-ph.HE].

\bibitem{Drury:2012md}
  L.~O.~Drury,
  arXiv:1203.3681 [astro-ph.HE].

\bibitem{Drury:1999}
  L.~O.~Drury, J.~P.~Meyer and D.~C.~Ellison,
  arXiv:astro-ph/9905008.

\bibitem{Clay:2010id}
  R.~W.~Clay, B.~J.~Whelan and P.~G.~Edwards,
  arXiv:1001.0813 [astro-ph.HE].

\bibitem{Taylor:2009we}
  A.~M.~Taylor, J.~A.~Hinton, P.~Blasi and M.~Ave,
  Phys.\ Rev.\ Lett.\  {\bf 103} (2009) 051102
  [arXiv:0904.3903 [astro-ph.HE]].

\bibitem{Abdo:2010nz}
  A.~A.~Abdo {\it et al.}  [Fermi-LAT Collaboration],
  Phys.\ Rev.\ Lett.\  {\bf 104} (2010) 101101
  [arXiv:1002.3603 [astro-ph.HE]].

\bibitem{Berezinsky:2010xa}
  V.~Berezinsky, A.~Gazizov, M.~Kachelriess and S.~Ostapchenko,
  Phys.\ Lett.\ B {\bf 695} (2011) 13
  [arXiv:1003.1496 [astro-ph.HE]].

\bibitem{Ahlers:2010fw}
  M.~Ahlers, L.~A.~Anchordoqui, M.~C.~Gonzalez-Garcia, F.~Halzen and S.~Sarkar,
  Astropart.\ Phys.\  {\bf 34} (2010) 106
  [arXiv:1005.2620 [astro-ph.HE]].

\bibitem{Abbasi:2011ji}
  R.~Abbasi {\it et al.}  [IceCube Collaboration],
  Phys.\ Rev.\ D {\bf 83} (2011) 092003
   [Erratum-ibid.\ D {\bf 84} (2011) 079902]
  [arXiv:1103.4250 [astro-ph.CO]].

\bibitem{Kronberg:1976} 
  P.~P.~Kronberg and M.~Simard-Normandin,
  Nature {\bf 263} (1976) 653

\bibitem{Kronberg:1982} 
  P.~P.~Kronberg and J.~J.~Perry,
  Astrophys.\ J.\ {\bf 263} (1982) 518

\bibitem{Blasi:1999hu}
  P.~Blasi, S.~Burles and A.~V.~Olinto,
  Astrophys.\ J.\  {\bf 514} (1999) L79
  [astro-ph/9812487].

\bibitem{Ahlers:2012az}
  M.~Ahlers, L.~A.~Anchordoqui and A.~M.~Taylor,
  Phys.\ Rev.\ D {\bf 87} (2013) 023004
  [arXiv:1209.5427 [astro-ph.HE]].

\bibitem{Takami:2012uw}
  H.~Takami, S.~Inoue and T.~Yamamoto,
  Astropart.\ Phys.\  {\bf 35} (2012) 767
  [arXiv:1202.2874 [astro-ph.HE]].

\bibitem{Kashti:2008bw}
  T.~Kashti and E.~Waxman,
  JCAP {\bf 0805} (2008) 006
  [arXiv:0801.4516 [astro-ph]].

\end{thebibliography}
\end{document}